\documentclass[11pt, letterpaper]{article}
\usepackage[dvipdfmx]{graphicx}
\usepackage{amsmath}
\usepackage{newtxtext}
\usepackage{authblk}
\usepackage{url}

\title{Concurrent Bursty Behavior of Social Sensors in Sporting Events}
\author{Yuki Takeichi}
\author{Kazutoshi Sasahara\thanks{Correspondence should be addressed to K.S. (sasahara@nagoya-u.jp)}}
\author{Reji Suzuki}
\author{Takaya Arita}
\affil{Department of Complex Systems Science, Nagoya University, \\Furo-cho, Chikusa-ku, Nagoya 458-8601, Japan}
\date{}

\begin{document}
\maketitle

\begin{abstract}
The advent of social media expands our ability to transmit information and connect with others instantly, which enables us to behave as ``social sensors.'' 
Here, we studied concurrent bursty behavior of Twitter users during major sporting events to determine their function as social sensors. 
We show that the degree of concurrent bursts in tweets (posts) and retweets (re-posts) works as a strong indicator of winning or losing a game. 
More specifically, our simple tweet analysis of Japanese professional baseball games in 2013 revealed that social sensors can immediately react to positive and negative events through bursts of tweets, but that positive events are more likely to induce a subsequent burst of retweets.
We confirm that these findings also hold true for tweets related to Major League Baseball games in 2015. 
Furthermore, we demonstrate active interactions among social sensors by constructing retweet networks during a baseball game. 
The resulting networks commonly exhibited user clusters depending on the baseball team, with a scale-free connectedness that is indicative of a substantial difference in user popularity as an information source.
While previous studies have mainly focused on bursts of tweets as a simple indicator of a real-world event, the temporal correlation between tweets and retweets implies unique aspects of social sensors, offering new insights into human behavior in a highly connected world.
\end{abstract}
\newpage

\section{Introduction}
Social media is an increasingly popular communication tool by which people have massive social interactions in cyberspace \cite{Miller2011}. 
These interactions can have a significant effect beyond cyberspace, with real world consequences. 
A well-known example is that social media helped Arab Spring activists spread and share information, playing a key role in the ensuing revolutionary social movements \cite{GonzalezBailon:2011kz}.
As in this case, social media can interface between cyberspace and the physical world by globally connecting people and information in nontrivial ways, thereby leading to novel collective phenomena. 
The quantitative understanding of such collective phenomena is a central issue in the emerging field of computational social science.
 
Many social media studies have been conducted using Twitter, a popular social media that allows users to read, post, and forward a short text message of 140 characters or less (called tweets).
These studies have focused on the characteristics and effects of Twitter, such as the structural properties of user networks \cite{Kwak:2010wy, Bollen:2011jo}, the nature of online social interactions \cite{Grabowicz:2012kt, Conover:2012cx} and information diffusion \cite{Romero:2011bg, Weng:2012dd}, collective attention \cite{Lehmann:2012, Sasahara:2013eu} and collective mood \cite{Golder:2011cy, Dodds:2011kk}, users' behavior related to particular real-world events \cite{Sakaki:2010, BorgeHolthoefer:2011jt}, and the prediction of the stock markets \cite{Bollen:2011hu}.
 
\begin{figure}[t]
\begin{center}
\includegraphics[width=80mm]{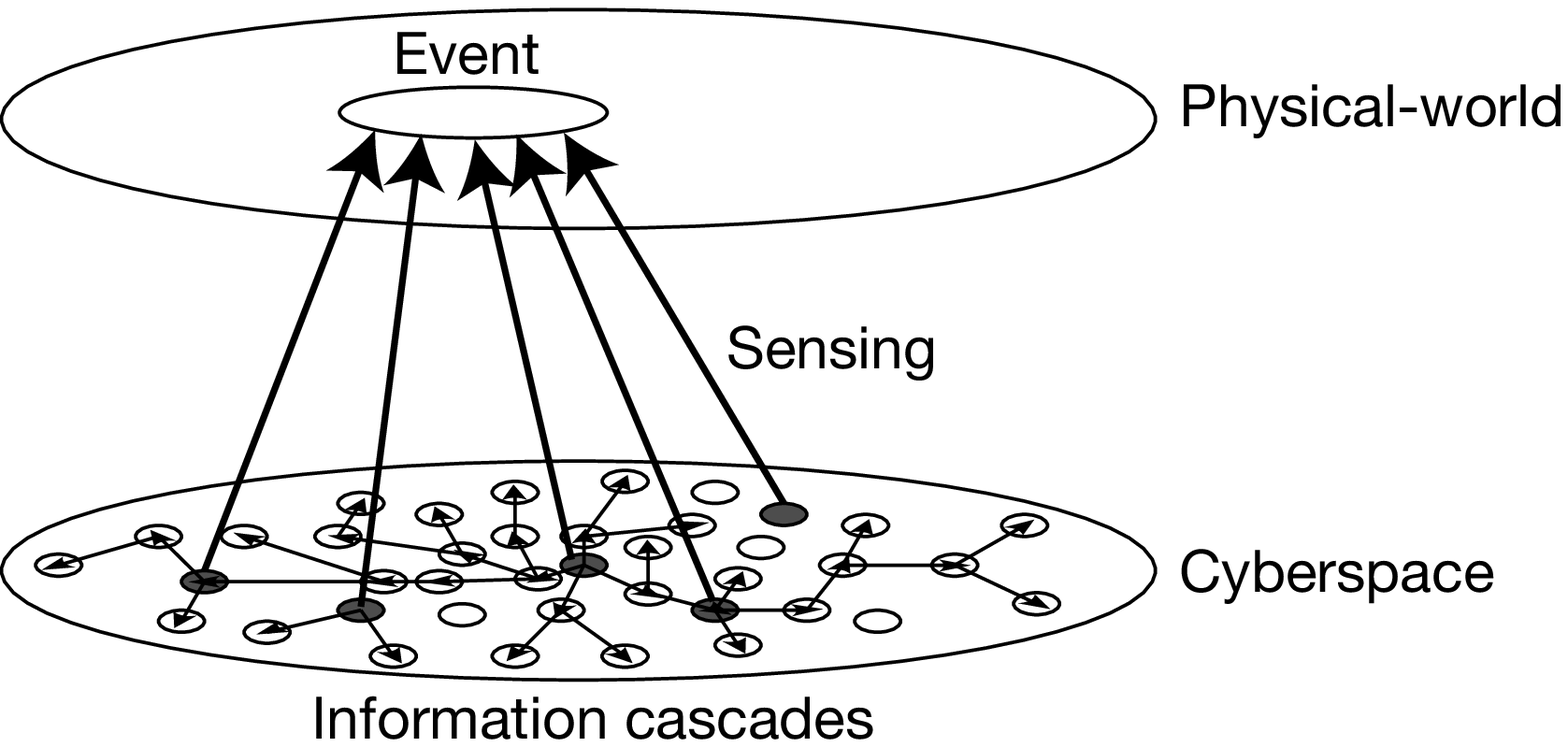}
\end{center}
\caption{Schematic illustration of a network of social sensors. Nodes in cyberspace represent social sensors (Twitter users). The thick arrows represent social sensors (grey nodes) sensing a real-world event, and thin arrows represent the corresponding information cascades by means of tweet and retweet.}
\label{fig:infocascade}
\end{figure}

In this paper, we focused on Twitter as a network of social sensors to investigate, a novel collective phenomenon empowered by social media.
Figure \ref{fig:infocascade} shows a schematic illustration of how social sensors work, in which Twitter users actively sense real-world events and spontaneously mention these events by posting tweets, which immediately spread over user networks in cyberspace. 
Such information cascades can be amplified by chains of retweets (re-posted tweets) from other users or followers.
Consequently, Twitter as a whole can behave like a network of social sensors, exhibiting distinct collective dynamics linked with target events.

Similar ideas have been tested in several different settings, most of which are in the context of the real-world event detection on Twitter.
For example, Sakaki and Matsuo monitored earthquake-related tweets and trained a statistical learning model with these data; they were successful (96$\%$ accuracy) in detecting earthquake events of the Japan Meteorological Agency of a seismic intensity scale three or more \cite{Sakaki:2010}.
Social sensors under emergency situations such as large earthquakes and Tsunamis were studied to demonstrate distinct retweet interactions \cite{Sasahara:2013eu}.  
Twitter data during sporting events were also analyzed in a variety of settings.
For example, Zhao et al. studied Twitter for real-time event detection during US National Football League (NFL) games and reported a detection accuracy of 90$\%$ in the most successful case \cite{Zhao:2011}.
Other studies developed methods for event detection from bursts of tweets related to football games by using a keyword frequency approach \cite{Lanagan2011, Corney:2014hi} and tweets about Olympic Games by using a non-negative matrix factorization approach \cite{Panisson2014}.

These studies share the hypothesis that Twitter is a mirror of reality and mainly focus on either bursts of tweets or retweets to identify spontaneous reactions of people to events in the physical world.
However, little is known about the more unique nature of social sensors that cannot be explained solely by these bursts of tweets or retweets.
Tweets and retweets, by nature, convey different kinds of information: tweets are more linked with what users want to say about real-world events, whereas retweets are more linked with what users are aware of in cyberspace.
Thus, the concurrent bursts of tweets and retweets would be a novel indicator of collective behavior.
The objective of this study was to determine the function of these concurrent bursts in social sensors.

\section{Materials and Methods}
\subsection{Dataset}
We targeted major sporting events for the study of social sensors.
This is because, as shown by the previous studies, natural disasters and major sporting events tend to strongly attract people's attention, which gives rise to a large volume of tweets and retweets. 
While natural disasters are largely unpredictable events, sporting events are scheduled and therefore allow data to be collected systematically.
Therefore, major sporting events were suitable targets for our aim.

Using the Twitter Search API ({\small \url{https://dev.twitter.com/docs/api/}}), which allows 180 queries per 15-min window, we compiled a dataset of tweets related to Japan's 2013 Nippon Professional Baseball (NPB) games, including at least one hashtag of NPB teams such as $\#${\tt giants} (Yomiuri Giants) and $\#${\tt rakuteneagles} (Tohoku Rakuten Golden Eagles).
These hashtags were selected by reference to a hashtag cloud site ({\small \url{ http://hashtagcloud.net}}). 
This hashtag-based crawling with multiple crawlers allowed us to obtain the nearly-complete data regarding these sporting events: 528,501 tweets surrounding 19 baseball games from the Climax Series (the annual playoff series) and from the Japan Series (the annual championship series) in the 2013 NPB. 
We also collected tweets related to Major League Baseball (MLB) games in 2015, including at least one hashtag of the MLB teams such as $\#${\tt Yankees} and $\#${\tt BlueJays}. 
The hashtags were selected by reference to Official Twitter account of the MLB ({\small \url{https://twitter.com/mlb}}).
We sampled 730,142 tweets from 17 games of New York Yankees from September 11 to 27, 2015.
The NPB complete dataset was used to address Twitter as a social sensor network and the MLB sampling dataset was used to validate the results of the former analysis.
The datasets are publicly available ({\small \url{http://dx.doi.org/10.7910/DVN/42V7E0}}).

\subsection{Correlation Between Tweet and Retweet Burst Trains}
Burst-like increases in tweets may arise when an event happens in the physical world, and that is what many previous researches on social sensors have shown.
However, in such cases, the reaction is not limited to tweets alone.
According to our observations, bursts of retweets often follow those of tweets when positive events happen in the physical world. 
If we assume tweet behaviors during a two-team sport, concurrent bursts of tweets and retweets would be repeatedly generated by the fans of the winning team and as a result tweet and retweet burst trains would be similar to each other with a little time lag.
With this point in mind, one of the easiest ways to measure the similarity of tweet and retweet burst trains is to use a cross-correlation function \cite{Box2015}. 
Suppose $x_i$ is a tweet count series and $y_i$ is a retweet count series, where $i=1$, $\cdots$, $N$. 
The cross-correlation function is defined as follows:
\begin{eqnarray*}
\label{equ:cross_corr}
r_{xy}(\tau) &=& \frac{1}{N-\tau}\sum_{i=1}^{N-\tau}\Bigl(\frac{x_i-\overline{x}}{\sigma_x}\Bigr)\Bigl(\frac{y_{i+\tau}-\overline{y}}{\sigma_y}\Bigr)
\end{eqnarray*}
where $\tau$ is the time lag, and $\overline{x}$ and $\sigma_{x}$ denote mean and variance, respectively. 
Its value ranges from -1 for complete inverse correlation to +1 for complete direct correlation. 
If $x_i$ and $y_i$ are not correlated, its value becomes around zero.
In this study, $x_i$ and $y_i$ were counted by 10 sec for the NPB complete dataset and those were counted by 60 sec for the MLB sampling dataset.
We set the maximum time lag 300 sec and adopted the maximum of $r_{xy}(\tau)$ as a measure of correlation between the tweet and retweet count series, denoted by $r_\text{max}$.
For statistical comparison, Fisher $z$-transformation \cite{Fisher1925} was applied to the resulting $r_\text{max}$ value to convert to the normally distributed value $R_{max}$. 
Thus, the greater $R_\text{max}$ indicates that tweet and retweet concurrent bursts highly correlate with each other.

\subsection{Construction of Retweet Networks}
The interactions of social sensors linked with major sporting events are examined using networks \cite{Newman:2010wp}. 
Using retweet data, we construct a retweet network as previously reported \cite{Sasahara:2013eu}. 
In the retweet network, each node represents a Twitter user and a directed edge is attached from user $B$ to user $A$, if user $B$ retweets a tweet originally posted by user $A$. 
If there is a user $C$'s retweet ``RT @user $B$ ... RT @user $A$ ...,'' then links are made $C \to B$  and $C \to A$.
In this network, retweet interactions among social sensors are represented and influential users (also known as hub users) whose tweets are preferentially retweeted by many users are represented as nodes with many incoming edges (in-degrees).

The resulting retweet networks are visualized in a force-directed layout algorithm in Gephi ({\small \url{https://gephi.org}}), so that users who retweet more frequently (i.e., more connections) can be placed closer together.
The size of nodes is proportional to the logarithm of the number of in-degrees.
In addition, cumulative in-degree distributions ($P_\text{cum}(k)=\sum_{k'}^{\infty}P(k')$) are calculated from retweet networks to access their structural properties.

\section{Results}
\subsection{Tweet and Retweet Bursts: An Example}
\begin{figure}[b]
\begin{center}
\includegraphics[width=120mm]{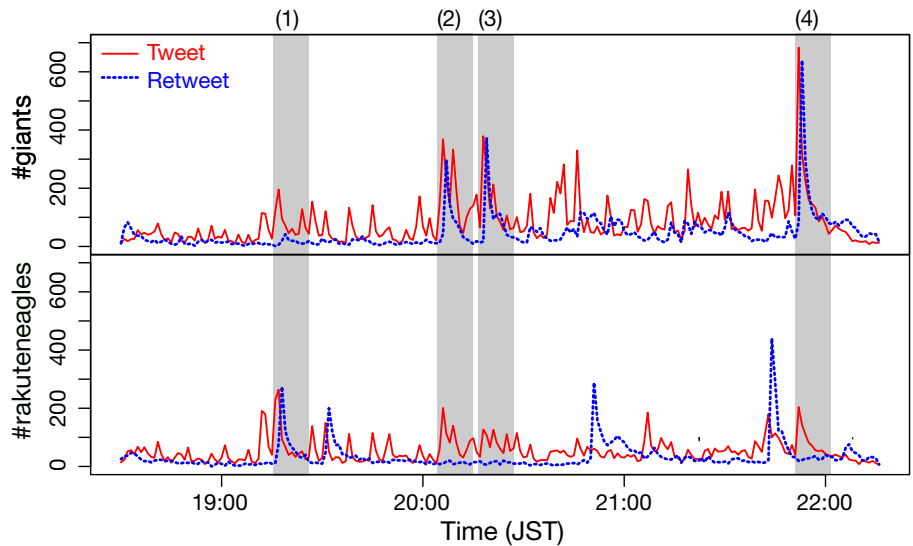}
\end{center}
\caption{Example of tweet and retweet time series (counts per minute) for the sixth round in the 2013 Japan Series. Red lines denote tweets and blue dashed lines denote retweets. The upper panel shows tweets for the Giants ($\#${\tt giants}) and the lower panel for the Eagles ($\#${\tt rakuteneagles}). See the main text for event (1)-(4). }
\label{fig:bbts}
\end{figure}

Figure \ref{fig:bbts} shows an example of the tweet and retweet dynamics during a baseball game, the 6th round in the 2013 Japan Series, in which the Yomiuri Giants beat the Tohoku Rakuten Golden Eagles by a score of 4-2.
We see many sudden increases of tweet and retweet counts for both teams, which are seemingly random spikes.
However, we noticed special cases where the bursts of tweets and those of retweets simultaneously occurred, and each of these cases corresponded to the following events, respectively:
\smallskip
\begin{enumerate}
\renewcommand{\labelenumi}{(\arabic{enumi})}
\item The Eagles scored twice.
\item The Giants scored third and turned the game around.
\item The Giants added another run.
\item The Giants won the game.
\label{tb:}
\end{enumerate}
\smallskip
As shown in Fig. \ref{fig:bbts}, the concurrent bursts of tweets and retweets were generated more frequently in the context of the Giants (the winning team) than the Eagles (the losing team).
Once a particular event happens during a game, users spontaneously post a scream of delight from the winning side and one of disappointment from the losing side.
For example, during event (3), positive tweets such as ``Oh goody!'' and ``Go-ahead homer!'' were posted with $\#${\tt giants}, whereas negative tweets such as ``Oh, no...'' and ``Disaster!'' were posted with $\#${\tt rakuteneagles}.
Without such events in a game, there was no strong bias against a tweet's polarity, positive or negative. 

This example shows that social sensors can immediately show reactions to a positive and a negative event by a burst of tweets; however, a positive event is more likely to induce a subsequent burst of retweets.
Therefore, we assume that a correlation between tweet and retweet time series would work as a measure of collective positive reactions of social sensors, which may eventually correlate to the result of a game.

\subsection{Tweet and Retweet Bursts During Games in the NPB}
\begin{figure}[h]
\begin{center}
\includegraphics[width=80mm]{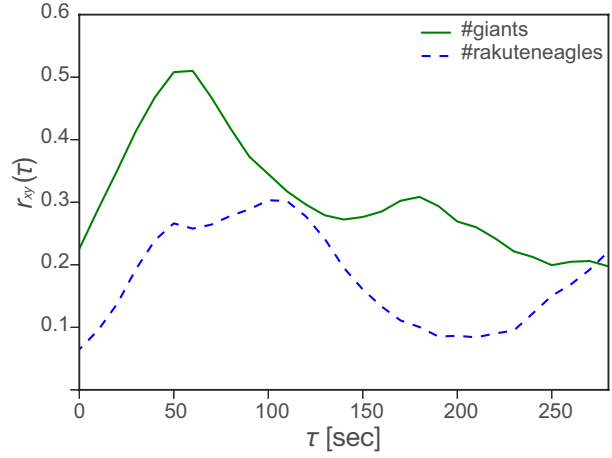}
\end{center}
\caption{Example of the correlation between tweet and retweet time series ($r_{xy}(\tau)$) for the six round in the 2013 Japan Series (cf. Fig. 2).}
\label{fig:r}
\end{figure}

We study the above-mentioned hypothesis using the NPB dataset.
To this end, we computed and compared $R_\text{max}$ for tweet and retweet time series, as defined in the Methods section, in 19 games from the Japan Series and the Climax Series for the Central and Pacific Leagues. 
Figure \ref{fig:r} shows an example of the correlation function ($r_{xy}(\tau)$) between tweet and retweet count series as a function of the time lag $\tau$ for the sixth round in the 2013 Japan Series, in which $r_{xy}$ reached the maximum at $\tau = 60$ sec for the Giants and at $\tau = 100$ sec for the Eagles.

\begin{figure}[h]
\begin{center}
\includegraphics[width=125mm]{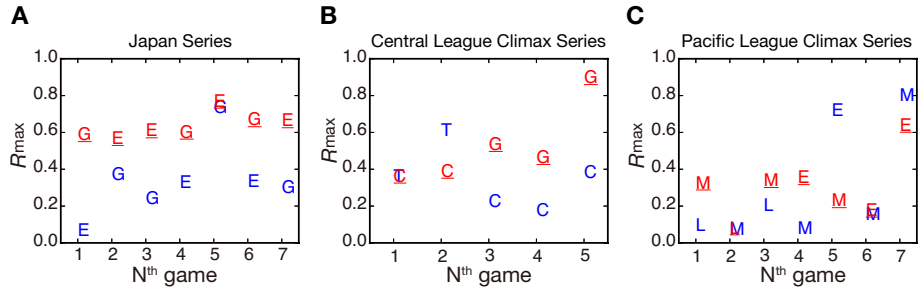}
\end{center}
\caption{$R_\text{max}$ between tweet and retweet time series for the 2013 Japan Series (A) and the 2013 Climax Series for the Central (B) and Pacific (C) Leagues. Red letters with an underline denote the winning team and blue letters denote the losing team. G: Yomiuri Giants, E: Tohoku Rakuten Golden Eagles, T: Hanshin Tigers, C: Hiroshima Toyo Carp, M: Chiba Lotte Marines, L: Saitama Seibu Lions.}
\label{fig:rmax}
\end{figure}

Figure \ref{fig:rmax} (left) shows the values of $R_\text{max}$ in tweet and retweet time series for the Giants (G) and the Eagles (E) across seven games in the Japan Series. 
In this figure, we can confirm that the winning teams have $R_\text{max}$ greater than that of the losing team in all games.
Moreover, two interesting features are shown in Fig. \ref{fig:rmax} (left): in the first round, $R_\text{max}$ for the Eagles was considerably smaller than that of the Giants, because the Eagles created scoring opportunities many times but failed to score a run; in the fifth round, both teams showed an equivalent $R_\text{max}$ value, because it was a closer game.
These results seem reasonable because a greater $R_\text{max}$ value is associated with positive events such as a base hit or a home run.

We then examined whether this property holds for other baseball games in the Climax Series. 
Figure \ref{fig:rmax} B and C reveal that this property holds true, except in the case of three games: the second round in the Central League Climax Series and the fifth and seventh rounds in the Pacific League Climax Series.
These exceptions were attributed to the non-stationary nature of tweet and retweet time series.
In two of these exceptions, the fans of a losing team generated a single intense concurrent burst of tweets and retweets when a scoring event happened in the late inning of the game. 
The other exception was based on an extraordinary number of retweets about the Eagles' victory in the Climax Series, which lowered the $R_\text{max}$ for the Eagles to below that of the losing team.
In principle, $R_\text{max}$ cannot be applied to a non-stationary time series; therefore both cases are out of the application range.
Overall, $R_\text{max}$ worked as a good indicator of the baseball game's results in 16 out of 19 games.
We also computed the time lag from the NPB dataset and the average time lag was 137$\pm$87 sec, at which correlation between tweet and retweet burst trains becomes maximum. 
There was not a significant difference in the time lag at $R_{max}$ between the winning team and the losing team (independent-samples $t$-test, $n$=38, $P$=0.059).

\begin{figure}[h]
\begin{center}
\includegraphics[width=60mm]{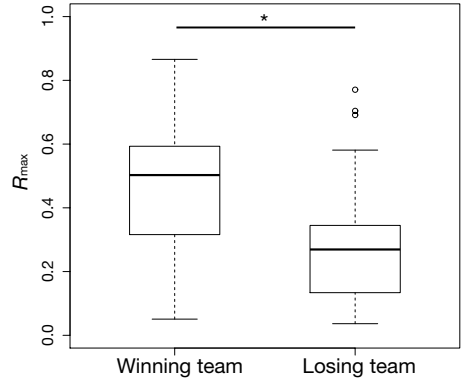}
\end{center}
\caption{Boxplots of $R_\text{max}$ in the winning team group and the losing team group in the 2013 Japan Series and the 2013 Climax Series for the Central and Pacific Leagues, with a significant difference between two groups.}
\label{fig:t-test}
\end{figure}

In Fig. \ref{fig:t-test}, we classified the computed $R_\text{max}$ values into two groups---one is the winning team group and the other the losing team group---and compared their means statistically.
The analysis identified a significant difference between the two groups (independent-samples $t$-test, $n=38$, $P < 0.05$), suggesting that greater $R_\text{max}$ values are related to winning games. 
Our hypothesis described above has now been statistically confirmed in the NPB dataset.

\subsection{Tweet--retweet Concurrency and Positive Events}
Here, we examined how social sensors reacted to positive events in the NPB baseball games.
We computed the relative occurrence frequency of ten baseball terms such as ``hit'' and ``homer'' ($r_\text{posi}$), as probes of positive events, from all of the baseball data. 
As a result, $r_\text{posi}$ is 0.07 $\pm$ 0.03 for tweets and is 0.28 $\pm $ 0.18 for retweets, indicating that retweets are more biased toward positive information than tweets (independent-samples $t$-test, $n$=38, $P < 0.001$). 
One expected result was that $r_\text{posi}$ for retweets would be higher in the winning team than in the losing team since retweets are used to convey positive information in a baseball game. 
Such correlation, however, was not confirmed (independent-samples $t$-test, $n=38$, $P=0.096$); in fact, $r_\text{posi}$ for retweets was higher in the losing team than in the winning team in 9 out of 19 games. 
These additional findings indicate that the number of positive tweets is not simply associated with wins or loses and that the timing or concurrency of tweet and retweet spikes ($R_\text{max}$) are more indicative of positive outcomes of sporting events.

\subsection{Tweet and Retweet Bursts During Games in the MLB}
\begin{figure}[b]
\begin{center}
\includegraphics[width=120mm]{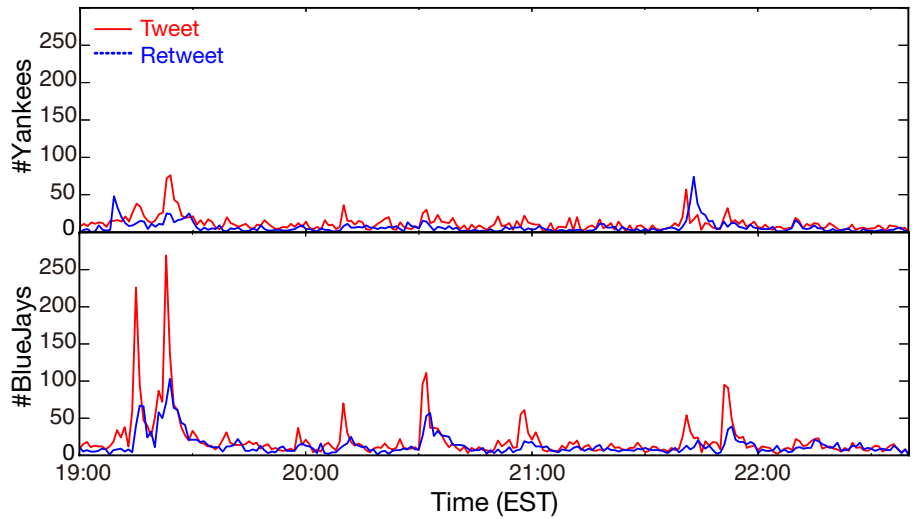}
\end{center}
\caption{Example of tweet and retweet time series (counts per minute) for the Yankees vs. Blue Jays game on September 11, 2015. Red lines denote tweets and blue dashed lines denote retweets. The upper panel shows tweets for the Yankees ($\#${\tt Yankees}) and the lower panel for the Blue Jays ($\#${\tt BlueJays}).}
\label{fig:tsmlb}
\end{figure}

There is potential concern that the above finding would be an artifact caused by the different Twitter usage or custom of Japanese users.
To confirm that this is not the case, we analyzed tweets sampled during New York Yankees games from September 11 to 27, 2015 ($n=17$), mostly posted by English-speaking users, with the same setting.
Figure \ref{fig:tsmlb} shows an example of tweet and retweet series counted by 60 sec for the first game in the above period, in which the Blue Jays beat the Yankees by a score of 11-5.
In this figure, concurrent spikes of tweets and retweets were associated with chances to score or scoring events, which is similar to Fig. \ref{fig:bbts}.
The resulting $R_\text{max}$ values for the MLB dataset in Fig. 7A show that the winning team had $R_\text{max}$ values greater than those of the losing team in 15 out of 17 games.
The two exceptions were seemingly due to the closeness of scores in the game.
There is a significant statistical difference in $R_{max}$ values between the winning team and the losing team values in Fig. 7B (independent-samples $t$-test, $n=34$, $P < 0.01$).
These results support our findings holding true across cultures, in that the concurrent bursts of tweets and retweets we observed are not, in fact, coincidental.

\subsection{Retweet Interactions Among Social Sensors}
To examine active interactions between social sensors, we constructed retweet networks related to different events in the sixth round in the 2013 Japan Series using a combined data set of tweets with $\#${\tt giants} and those with $\#${\tt rakuteneagles}.
As mentioned before, nodes represent Twitter users, who are fans of either team or baseball fans in general, and directed links represent official retweets between them. 

In Fig. \ref{fig:rtnet}, the retweet network (A) corresponds to event (1) where the Eagles got two runs in the second inning, and the network (B) corresponds to events (2) and (3) where the Giants turned the game around.
These networks are composed of two main sub-networks, one is a cluster of the Giants fans (green) and the other is a cluster of the Eagles fans (blue). 
While a large amount of retweets were transferred within the same sub-networks (i.e., the fans of the same team), there were much fewer retweets between the different sub-networks.
Interestingly, there were a few retweets with both hashtags.
Moreover, the Giants cluster involves several hub users (large nodes) who are preferentially retweeted by many users, whereas only a single hub user existed in the Eagles cluster.
It turned out that these hub users are either the official account for the teams or enthusiastic baseball fans.

The bottom panels in Fig. \ref{fig:rtnet} show the cumulative in-degree distributions of the retweet networks (A) and (B), respectively.
Both of the distributions exhibit a scale-free property 
Furthermore, the tails tended to shift to the right (i.e., greater $k$) on the winning side; that is, the tail is much longer in the Eagles cluster than the Giants cluster in (A), while the situation is opposite in (B).

These structural properties provide additional clues on how social sensors act, react, and interact to generate collective busty behavior during a sporting event.
First, the scale-free property of retweet networks is indicative of a substantial difference in the popularity of social sensors as an information source for retweets.
Second, the existence of two main sub-networks suggests that social sensors self-organized topic-based groups, in which they had a sense of belonging in their groups by using the same hashtag.

\section{Discussion}
We have demonstrated that social sensors respond preferentially to positive events in sporting events by generating concurrent bursts of tweets and retweets, the degree of which can be interpreted as a strong indicator of winning or losing a game.
We think that such concurrent reaction occurs in a wide variety of settings but it is often weak or one-time occurrences, neither of which is a condition that fits our approach. 
Thus, we used major sporting events as ideal exemplars to illustrate the concurrent bursty behavior of social sensors that previous research has not addressed.
A burst of tweets reflects a fast process where social sensors respond reflexively to real-world events, whereas that of retweets reflects a slower process where social sensors become aware and react selectively to the information posted about these events in cyberspace.
As the latter process requires more attention and is highly context dependent, concurrent bursts of tweets and retweets are seemingly unlikely but possible during positive real-world events, as we have demonstrated.
As seen in Fig. 8, there are a few hub users (or influentials) who can cause data bias, therefore the amount of tweets (or retweets) cannot be a good measure for social sensors but the degree of the tweet--retweet concurrency can be a much robust measure. 
By incorporating this nature of concurrency with the conventional measures, we can develop a more accurate, reliable indicator of positive real-world events; otherwise, every single measure alone cannot work.
Several exceptions observed in the baseball data suggest that the tweet--retweet concurrency is only one aspect of social sensors and that much remains to be discovered. 
Therefore, exploring real-world events by focusing on different features is indispensable for understanding the true complexity of social sensors.
An extension of this study in this direction is also important for the development of an application of real-time social sensing, using humans as sensors, in the social media system of the future.

Our findings, however, do not necessarily hold true in other sporting events because different sports have different scoring dynamics \cite{Merritt:2014jv}.
For example, two-team sports such as baseball and football have detailed rules with a scoring mechanism that can prompt fans to be more aware of a game's progress. 
This situation tends to elicit spontaneous, polarized tweet and retweet reactions to chances to score and scoring events among fans of different teams. 
In contrast, in multi-team sports like car racing, the rules are much simpler and there is no scoring mechanism, which may deprive fans of a chance to react to the progress of a race. 
In this situation, tweet and retweet reactions occur in a different fashion than with two-team sports.
Furthermore, there are potential disadvantages of this method. 
As mentioned earlier, the long, stationary time series is necessary for the accurate estimation of $R_\text{max}$. 
This is because the cross correlation function is a linear measure and it can poorly capture correlations between nonlinear signals; in such a case more advanced but perhaps more computationally expensive measures need to be employed. 
Our approach cannot work in non-popular sporting events, because people hardly tweet for such events and hence the amount of available tweets is not enough for analysis.
Although several limitations are recognized, we think that the temporal correlation between tweets and retweets is a good measure to explore social sensors, and $R_\text{max}$ can be applied to a wider class of sporting events and probably other social events, such as presidential debates between two candidates.

Furthermore, the retweet networks for the baseball games exhibited a scale-free property of user popularity, with hub sensors (or influentials) who contribute to cascades of retweets, as with other retweet networks for meme diffusion \cite{Weng:2012dd} and for collective attention \cite{Sasahara:2013eu}.
In addition, these retweet networks had sub-networks depending on the baseball teams, as with user networks for online political activity \cite{Conover:2012cx}.
These sub-networks are interpreted as ``topic-based groups'' \cite{Ren2006}, in which people feel attached to the group or loosely connected to one another, by using the same hashtag.
The common structural features of social sensor groups indicate the possibility of the same underlying design principle.
To assess the generality of these results, further investigations are necessary using a wide variety of social events across various kinds of social media.

In conclusion, our simple analysis provides evidence that Twitter is a network of social sensors in that it allows people to immediately react to real-time events by tweeting and it is active in that people selectively retweet favorite posts, thereby yielding the spontaneous concurrent bursts of tweets and retweets that spread over scale-free user networks.
Contrary to the well-tested analogy that ``Twitter is a mirror of reality,'' the results of this study imply the more unique aspects of social sensors, few of which have been quantitatively addressed so far. 
The accumulation of case studies of this kind is fundamental for computational social science to understand the complexity of human behavior in a highly connected world.



\newpage

\begin{figure}[t]
\begin{center}
\includegraphics[width=120mm]{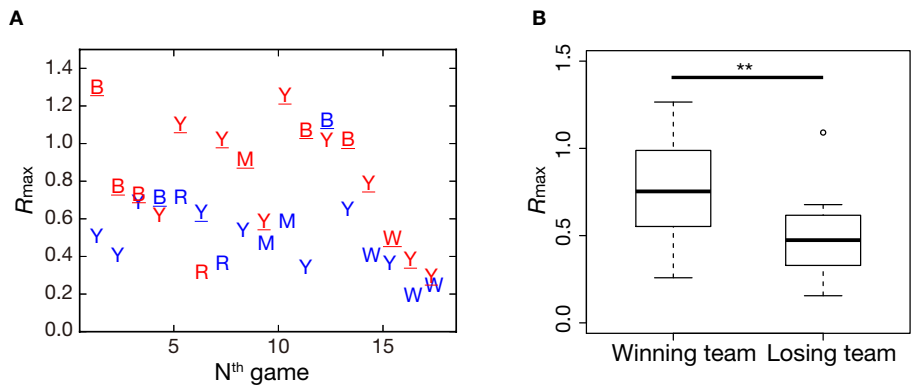}
\end{center}
\caption{$R_\text{max}$ values between tweet and retweet time series for the 2015 Major League Baseball (Yankees games from September 11 to 27). (A) $R_\text{max}$ values by games ($n=17$). Red letters with an underline denote the winning team and blue letters denote the losing team. Y: New York Yankees, B: Toronto Blue Jays, M: New York Mets, R: Tampa Bay Rays, W: Chicago White Sox. (B) Boxplots of $R_\text{max}$ in the winning team group and the losing team group, with a significant difference between two groups.}
\label{fig:rmaxmlb}
\end{figure}

\begin{figure}[t]
\begin{center}
\includegraphics[width=120mm]{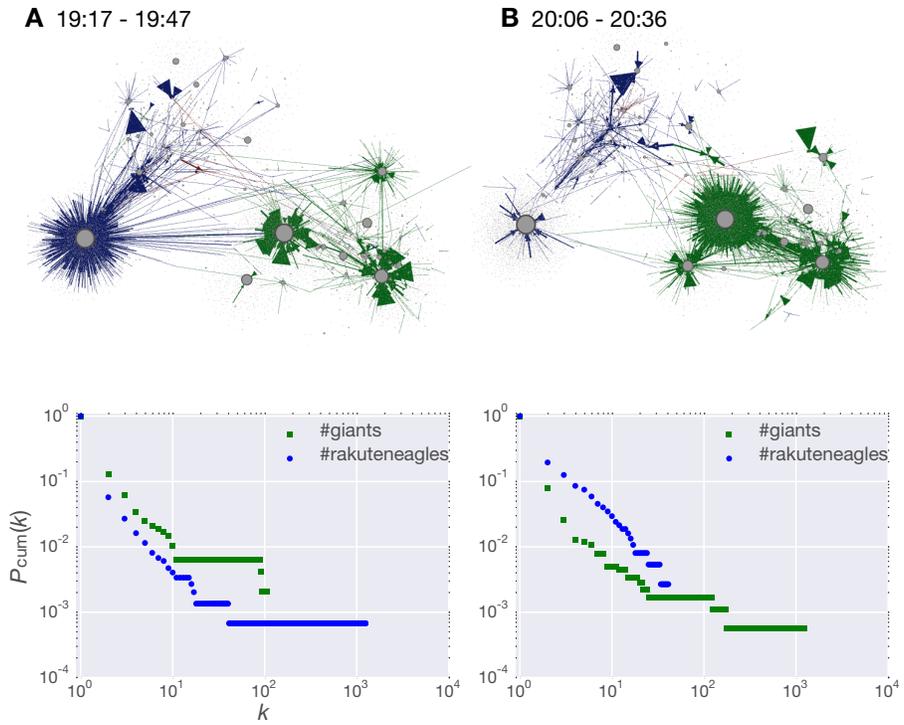}
\end{center}
\caption{Retweet networks and their cumulative in-degree distributions ($P_\text{cum}(k)$) in the sixth round of the 2013 Japan Series. The retweet network (A) consists of data generated during 30 min from 19:17, in which more retweets were generated with $\#${\tt rakuteneagles}. The retweet network (B) consists of data generated during 30 min from 20:16, in which more retweets were generated with $\#${\tt giants}. Green lines and circles denote $\#${\tt giants} and blue lines and circles denote $\#${\tt rakuteneagles}.}
\label{fig:rtnet}
\end{figure}

\end{document}